\documentclass[aps,prd,twocolumn,showpacs,amsmath,nofootinbib]{revtex4-1}

\usepackage{color}
\newcommand{\beqa}{\begin{eqnarray}}
\newcommand{\eeqa}{\end{eqnarray}}
\usepackage{float}
\usepackage{epsfig}
\usepackage{graphicx}
\usepackage{latexsym}
\usepackage{amsmath}
\usepackage{hyperref}
\usepackage{mathrsfs}
\usepackage{dcolumn}% Align table columns on decimal point
\usepackage{bm}%
\usepackage{multirow,booktabs}

\begin{document}

\title{Constraints on redshift-evolving Hubble constant models with early- and late-time diagnostics}

\author{Yupeng Yang}\email{ypyang@aliyun.com}
\affiliation{School of Physics and Physical Engineering, Qufu Normal University, Qufu, Shandong, 273165, China}

\begin{abstract}

Recent analyses, in which supernova data are binned in redshift and Hubble constant $H_0$ is fitted separately in each bin, reveal a systematic trend of 
$H_0$ with redshift, suggesting that $H_0$ may not be a true constant. In this work, we systematically investigate four cosmological models within a parameterized framework that allows for a redshift dependence of \(H_0\): three phenomenological models (\(\alpha\Lambda\)CDM, \(\alpha w\)CDM, and \(\alpha w_0w_a\)CDM), and the $J$CDM model motivated by big bang quantum cosmology. Using cosmic microwave background (CMB) data, baryon acoustic oscillations (BAO), cosmic chronometer (CC) measurements of the Hubble parameter, and three Type Ia supernova samples (PantheonPlus, DESY5, and Union3), we perform Markov chain Monte Carlo (MCMC) analyses to obtain posterior distributions of the model parameters. We also perform separate analyses using early-time and late-time data to assess how each observational epoch affects the model constraints.
For the three phenomenological models, the parameter \(\alpha\) describing the redshift evolution of \(H_0\) is consistent with zero within \(2\sigma\). The Hubble constant inferred from the \(\alpha\Lambda\)CDM model is consistent with CMB results, while the $J$CDM model provides excellent fits to late-time 
data and satisfies the Planck BAO constraint on $\Omega_mh^2$ for SNIa samples; however, its CMB-only fit is strongly inconsistent with Planck, 
and the opposite holds for $\Lambda$CDM. 
Using the Akaike Information Criterion (AIC) and the Bayesian Information Criterion (BIC) for model comparison, we find that the standard \(\Lambda\)CDM model remains the most favored by the data. Among the four models, the \(\alpha w_0w_a\)CDM model shows a marginal AIC advantage 
($\rm \Delta AIC \sim$ -2 to -13). However, this preference does not translate into a reduction of the Hubble tension, which is consistent with existing studies supporting dynamical dark energy. Moreover, none of the models can simultaneously satisfy both 
the early- and late-time constraints, and the combined datasets alone are insufficient to conclusively determine whether these parametrizations can resolve the tension.

\end{abstract}

\maketitle

%%%%%%%%%%%%%%%%%%%%%%%%%%%%%%% Introduction %%%%%%%%%%%%%%%%%%%%%%%%%%%%%%%%%%%%%%%%%
\section{Introduction}

Although the standard cosmological model, the $\Lambda$ cold dark matter ($\Lambda$CDM) model, has been successful in 
describing many cosmological observations, several tensions exist between the model and observations~\cite{DiValentino:2025otz,CosmoVerseNetwork:2025alb}. 
The ``Hubble tension", as a major cosmological tension, has attracted the most attention (see, e.g., Refs.~\cite{DiValentino:2021izs,Hu:2023jqc,Kamionkowski:2022pkx,
H0LiCOW:2019pvv} for a review). The Hubble tension refers to the tension between early- and late-time cosmology, manifested as the inconsistency between 
the local distance-ladder (LDL) measurement of $H_0$ (e.g., SH0ES) and the value inferred from Planck CMB data under the $\Lambda$CDM model. 
The Hubble constant ($H_0$), an important cosmological parameter, has been measured using many different astronomical observations. 
There are various ways to derive this parameter, mainly referring to the early and late Universe. 
For the early Universe, the method involves measuring the cosmic microwave background (CMB). Results 
from the Planck satellite show that the Hubble constant is $H_{0} = 67.4\pm{0.5}\rm km~s^{-1}~Mpc^{-1}$, 
which is model dependent and based on the $\Lambda$CDM model~\cite{2018Planck}. A model-independent method, based on the LDL, 
from the SH0ES collaboration gives the Hubble constant as $H_{0} = 73.04\pm{1.04}\rm km~s^{-1}~Mpc^{-1}$~\cite{2021A}. 
The discrepancy between the measured values reaches about $4.9\sigma$. Note that, while this tension was initially 
prompted by the $H_0$ estimates from LDL measurements versus the Planck-$\Lambda$CDM 
analysis of the CMB, it extends to analyses of models that include the BAO constraint $\Omega_mh^2$~\cite{Abchouyeh:2025ans_2026}. 
Many models have been proposed to resolve this tension, and this tension would imply the existence of new physics~\cite{Kamionkowski:2022pkx,Jia:2025prq,Rezaei:2024vtg,
2024hct..book..277M,Poulin_2019,Yang:2025vnm,Brito:2024bhh,Elizalde:2021kmo,Khurshudyan:2024gpn,Wu:2025vfs,Cai:2021weh,HU:2026syg,Dai:2026xau,Li:2026asg,Singh:2026bxh,Singh:2026ejb,Al-Omar:2026cjo,Maier:2026nha,Wang:2026sqy,Millano:2026hjk,LaPenna:2026avs,deCruzPerez:2025dni,Li:2025vqt,Sabogal:2026ipu,Wang:2026tgi,Shah:2026oxn,Schiavone:2026agq,Li:2025owk,Li:2026xaz,DiValentino:2019exe,Paliathanasis:2026ymi,vanderWesthuizen:2025iam,Paliathanasis:2025xxm,Abchouyeh:2025ans,Ren:2025jhe,Meng:2026lzz}.

Recently, a redshift evolution of the Hubble constant has been proposed based on related observations~\cite{Montani:2024ntj,Dainotti:2022bzg,Jia:2022ycc,Colgain:2022nlb,
Colgain:2022rxy,Jia:2024wix,Krishnan:2020obg,Krishnan:2020vaf,DeSimone:2024lvy,Lopez-Hernandez:2024osv}. 
Using type Ia supernova (SNIa) data, which refer to the late Universe, the authors of~\cite{Dainotti:2021pqg} found that 
the Hubble constant derived from different epochs differs and shows an evolutionary trend. Note that these analyses refer to a $\Lambda$CDM normalization, 
obtained by fitting $\Lambda$CDM separately to redshift-binned late-time SNIa data. 
The results of~\cite{Dainotti:2025qxz,Hu:2022kes,Dainotti:2022bzg,
Jia:2024wix,Dainotti:2025qxz}, including the CC and BAO data, 
confirmed this finding. The evolution of $H_0$ can be parameterized as 

\beqa
\mathcal{H}_0(z) = \frac{\widetilde{H_0}}{(1+z)^{\alpha}},
\label{eq:h0z}
\eeqa
with a constant $\alpha \gtrsim  0$. Note that $\alpha=0$ corresponds to $\Lambda$CDM model. 
For example, the fitting results from~\cite{Dainotti:2025qxz} show that $\alpha=0.010\pm{0.006}$ with 
$\widetilde{H_{0}} = 70.04\pm{0.14}\mathcal~\rm km~s^{-1}~Mpc^{-1}$ using the SNIa data~\footnote{This result can be found in Table~2 of~\cite{Dainotti:2025qxz}, 
where several fits corresponding to different data sets and models are presented. We have chosen one for illustration.}. 
In Refs.~\cite{Dainotti:2025qxz,Dainotti:2021pqg,Schiavone:2022wvq}, the authors proposed that the $f(R)$ model would produce this trend. The authors of~\cite{Dai:2026pvx} investigated 
a specific interacting model between dark matter and dark energy, and found that this interacting model 
can be rewritten as a form similar to Eq.~(\ref{eq:h0z}). In Ref.~\cite{Yang:2025oax}, the author examined 
the deviation of Hubble expansion from the $\Lambda$CDM model and found that a similar form can also be obtained for a specific choice. 
On the other hand, the authors of~\cite{Mo:2026dlk} argued that the evolutionary trend of the Hubble constant might arise from known parameter degeneracies. 
Although the underlying physics is currently unknown, the evolution trend 
of the Hubble constant provides a consistent potential solution to the Hubble tension. 
 
Recently, a $J$CDM big bang quantum cosmology model has been proposed~\cite{van_Putten_2025,van_Putten_2015}. In this model, the time-translation symmetry is violated 
on the Hubble time scale, leading to the presence of vacuum energy in the form of residual thermal energy. 
This dynamic dark energy component is characterized by the trace $J$ of the Schouten tensor as $J=R/6$, where $R$ is the scalar curvature of the background cosmology. 
In this model, the cosmological constant $\Lambda$ in the $\Lambda$CDM model is replaced by $J$, which is a dynamic quantity that 
evolves with the expansion of the Universe. The Hubble expansion rate in the $J$CDM model has a similar form to Eq.~(\ref{eq:h0z}) 
and has the potential to alleviate the Hubble tension. It is also worth noting that the $J$CDM model, originally developed for late-time cosmology~\cite{van_Putten_2015}, has recently been generalized to incorporate a breaking of time-reversal symmetry, which manifests as non-adiabaticity in the Hubble expansion. This extension permits a direct confrontation with Planck data within a fully nonlinear $w(a)$CDM analysis, 
while leaving the late-time predictions for the LDL essentially unchanged \cite{VanPutten:2026Jt}.

In this work, motivated by the evolution trend of the Hubble constant, we consider three parameterized models phenomenologically. 
These can be regarded as formal extensions of the most studied $\Lambda$CDM, $w$CDM, and $w_0 w_a$CDM models 
with an extra parameter $\alpha$. Using a combination of datasets from cosmic microwave background (CMB), 
baryon acoustic oscillations (BAO), cosmic chronometers (CC), and SNIa, we derive the posterior 
distribution of parameters in different models and investigate their ability to resolve the Hubble tension. 
We also conduct an analysis of the $J$CDM model. Moreover, we compare these four models using the 
Akaike Information Criterion (AIC) and the Bayesian Information Criterion (BIC). It is important to point out that the Hubble tension stems from the discrepancy between 
the values of $H_0$ derived from early-time and late-time probes. To investigate the impact of different cosmic epochs on the models, we perform separate model fits using early-time data (CMB) and late-time data (BAO+CC+SNIa). This approach enables us to diagnose whether any model alleviates the tension by modifying early- or late-universe physics, and at what cost to consistency with the other epoch.

This paper is organized as follows. In Sec. II we provide the Hubble expansion equations for these four cosmological models. 
We then describe the datasets and methods employed in our analysis in Sec. III. In Sec. IV, 
the constraints on the parameters and related discussion are presented. The conclusions are given in Sec. V.
%%%%%%%%%%%%%%%%%%%%%

\section{The Hubble expansion rate of three parameterized and $J$CDM models }

As mentioned in the previous section, motivated by the evolution trend of the Hubble constant, we here investigate three parameterized 
models phenomenologically. These three parametric models can be regarded as generalizations of the most studied model 
that incorporates Eq.~(\ref{eq:h0z}). The specific forms of the Hubble expansion rate for the three models are as follows, 

(i) $\alpha \Lambda$CDM model:
\beqa
H^2(z) = \frac{H_0^2}{(1+z)^{2\alpha}}\left[\Omega_{m}(1+z)^{3}+\Omega_{\Lambda}+\Omega_{r}(1+z)^4\right]
\eeqa  

(ii) $\alpha w$CDM model:

\beqa
H^2(z) = &&\frac{H_0^2}{(1+z)^{2\alpha}}\biggl[\Omega_{m}(1+z)^{3} \nonumber \\
&&+\Omega_{\rm DE}(1+z)^{3(1+w)}+\Omega_{r}(1+z)^4\biggr]
\eeqa

(iii) $\alpha w_0 w_a$CDM model:

\beqa
H^2(z) =&& \frac{H_0^2}{(1+z)^{2\alpha}}\biggl[\Omega_{m}(1+z)^{3} \nonumber \\
      &&+ \Omega_{\rm DE}(1+z)^{3(1+w_0+w_a)}\exp\left(-3w_a\frac{z}{1+z}\right) \nonumber \\
      && +\Omega_{r}(1+z)^4\biggr]
\eeqa
where $\Omega_m$, $\Omega_\Lambda$, $\Omega_{\rm DE}$, and \(\Omega_r\) are dimensionless parameters for matter, 
the cosmological constant, dynamical dark energy, and radiation at $z=0$, respectively. 
Note that we consider a flat Universe here. 

In addition to the above three parameterized models, we also consider the \(J\)CDM cosmological model, which has no 
additional parameters compared with the standard model and shows a similar form to those models. 
The Hubble expansion rate can be written as~\cite{van_Putten_2025,van_Putten_2026} 

\beqa
H^2(z)=\frac{H_0^2}{(1+z)^2} \left[1+\frac{6}{5}\Omega_{m}Z_{5}(z)+\Omega_{r}Z_{6}(z)\right],
\eeqa
here, $Z_{n}(z) = (1+z)^{n} - 1$, and we have set the curvature density $\Omega_k =0$, corresponding to the flat $J$CDM model. 
In Ref.~\cite{Wang:2026rkx}, the authors performed an analysis of the $J$CDM model using the datasets CMB+BAO+CC, 
which are the same as those used in our work. For the SNIa data, Ref.~\cite{Wang:2026rkx} used the Pantheon dataset, 
whereas we have extended the SNIa data to include the PantheonPlus, DESY5, and Union3 samples for our analysis. 

%%%%%%%%%%%%%%%%%%%%%%%%%%%%%%%%%%%%%%%%%%%%%%%%%%%%%%%%%%%%%%%%%%%%%%%%%%%%%%%%%%%%%%

\section{Analysis of Data and Related Formulas}

\subsection{Baryon acoustic oscillations}

Baryon acoustic oscillations (BAO) are the imprint of density waves generated by early cosmic baryon density perturbations. 
They can be regarded as a standard ruler for measuring the distances in the Universe. BAO measurements have already been 
obtained~\cite{2012JCAP...03..027G,Blake_2011,2010MNRAS.401.2148P,2011MNRAS.416.3017B,SDSS:2005xqv}. 
Recently, DESI conducted the most precise BAO measurements to date and successively 
released the associated results DR1~\cite{DESI:2024mwx} and DR2~\cite{DESI:2025zgx}. In this work, we will use the most recent results, 
namely the data from DR2. The description of the physical quantities related to BAO, 
the transverse comoving distance \(D_{M}(z)\) and the angular average distance $D_{\rm V}(z)$, is as follows.

For a homogeneous and isotropic Universe, the transverse comoving distance can be written as
\beqa
D_{M}(z)=\frac{c}{H_{0}\sqrt{|\Omega_{k}|} }{\rm sinn} \left [ \sqrt{|\Omega_{k}|} \int_{0}^{z}  \frac{dz^{' }}{H(z^{'})/H_{0}} \right ]
\eeqa
where $c$ denotes the speed of light, and $H(z)$ is the Hubble expansion rate. ${\rm sinn}(x) = {\rm sinh}(x), x, {\rm sin}(x)$ corresponds to an open, 
flat, and closed universe, respectively. The angular average distance $D_{\rm V}$ is 
then written as~\cite{Eisenstein_2005,Zhang:2025qbs}
\beqa
D_{V}(z) = [z D_{M}^2(z) D_{H}(z)]^{1/3},
\eeqa
where $D_{H}(z) = c / H(z)$. 

For the DESI BAO data, the samples include the Luminous Red Galaxy (LRG) sample with effective redshifts \(z_{\text{eff}} = 0.51\) 
and \(0.71\), the combined LRG + Emission Line Galaxy (ELG) sample, the ELG sample, and the Lyman-alpha 
Forest Quasar (Ly\(\alpha\) QSO) sample. For these samples, the standardized distance parameters \(D_M/r_d\) and \(D_H/r_d\) are considered, 
where $r_d$ is the acoustic horizon. The chi-square statistic for these samples can be expressed as~\cite{Kamionkowski:2022pkx}
\beqa
{\chi ^{2} _{1}}=\sum_{i}^{} \bigtriangleup D_{i}^{T}{\rm Cov}_{\rm BAO}^{-1}\bigtriangleup D_{i},
\eeqa
where
\beqa
D_{i}=\begin{bmatrix}D_{M}/r_{d}\\D_{H}/r_{d}\end{bmatrix},
\eeqa
and
$\Delta D_{i}=D_{i}^{\rm th}-D_{i}^{\rm data}$.

For the bright galaxy sample (BGS, \(z_{\text{eff}}=0.30\)) and the quasar sample (QSO, \(z_{\text{eff}}=1.49\)), 
the chi-square statistic can be expressed as

\beqa
\chi^2 _{2}= \sum_{k} \left( \frac{D_V^{\rm th}/r_d - D_V^{\text{obs}}/r_d}{\sigma_{D_V}} \right)^2.
\eeqa

For the total DESI BAO samples, the total $\chi^{2}$ is given by
\beqa
\chi^2 _{\rm BAO}=\chi^2 _{1}+\chi^2 _{2}
\eeqa

%%%%%%%%%%%%%%%%%%%%%%%%%%%%%%%%%%%%%
\subsection{Cosmic microwave background} 

The angular power spectrum of cosmic microwave background (CMB) is commonly used to constrain cosmological models. Alternatively, 
CMB distance priors can provide an efficient and widely adopted approach for parameter estimation~\cite{Chen_2019,Wang:2007mza,Zhai:2019nad}, 
as demonstrated in many studies (e.g., Refs.~\cite{Yang:2025vnm,Yang:2025oax,8ync-vrtz,Zhai:2018vmm,Li:2024hrv,Jia:2025prq,Rezaei:2024vtg,Sohail:2024oki}). 
In this work, we employ CMB distance priors, including the sound horizon encoded in the acoustic scale $l_a$, the shift parameter $R$, 
and the baryon density $\Omega_b h^2$, to derive robust constraints on the relevant parameters~\cite{Liu_2019, Xu_2016, Komatsu_2009, Yang:2025oax}. 
It should be noted that, while we use distance prior to obtain final constraints, 
a full analysis of the CMB angular power spectrum data would in principle yield more accurate results.

The acoustic scale and shift parameter are defined as~\cite{2018Planck}

\beqa
l_{a}\equiv  (1+z_{*})\frac{\pi D_{A}(z_{*})}{r_{s}(z_{*})}
\label{eq:cmb_la}
\eeqa

\beqa
R\equiv \sqrt{\Omega _{m}H_{0}^2} (1+z_{*})D_{A}(z_{*})
\label{eq:cmb_R}
\eeqa
where $z_{*}$ denotes the redshift at the epoch of photon decoupling, and \(r_{s}\) is the comoving sound horizon. We adopt an approximate form for \(z_{*}\) following~\cite{Hu_1996}\footnote{More accurate forms exist (see Ref.~\cite{Aizpuru:2021vhd}), but using them has only a slight impact on our final results.}:

\beqa
z_{*}=1048[1+0.00124(\Omega_{b}h^{2} )^{-0.738}][1+g_{1}(\Omega_{m}h^{2})^{g_{2}}]
\eeqa
with $h=H_0/100 \rm km~s^{-1}~Mpc^{-1}$ and 
\beqa
g_{1}&=&\frac{0.0783(\Omega_{b}h^{2})^{-0.238}}{1+39.5(\Omega_{b}h^{2})^{0.763}}\\
g_{2}&=&\frac{0.56}{1+21.1(\Omega_{b}h^{2})^{1.81}}.
\eeqa

The comoving sound horizon $r_{s}$ is given by
\beqa
r_{s}(z)=\frac{c}{H_{0}} \int_{0}^{1/(1+z)} \frac{da}{a^{2}h(a)\sqrt{3(1+\frac{3\Omega_{b}h^{2}}{4\Omega_{\gamma }h^{2}}a )}},
\eeqa
where $a=1/(1+z)$ and
\beqa
\frac{3}{4\Omega_{\gamma }h^{2}}=31500(T_{\rm CMB}/2.7K)^{-4}
\eeqa
with $T_{\rm CMB}$=2.7255K. The angular diameter distance $D_A$ can be written as
\beqa
D_{A}(z)=\frac{D_M(z)}{1+z}.
\eeqa

For the CMB data,  the chi-square statistic is expressed as
\beqa
\chi ^{2}_{\rm CMB}=\Delta X^{T}{\rm Cov}^{-1}_{\rm CMB}\Delta X,
\eeqa
where $\Delta X$=$X^{\rm th}- X^{\rm obs}$. The vector \( X^{\text{obs}} \) consists of three parameters: 
the shift parameter \( R \), the acoustic scale \( l_a \), and the baryon density parameter \( \Omega_b h^2 \). 
The observed values \( X^{\text{obs}} \) and the covariance matrix \( \mathbf{\rm Cov}_{\rm CMB} \) used in this study are 
taken from Planck 2018 as given in, e.g, Ref.~\cite{Chen_2019}.

%%%%%%%%%%%%%%%%%%%%%%%%%%%%%%%%%%%%%%%

\subsection{Cosmic chronometers}

We utilize cosmic chronometer (CC) data to obtain measurements of the cosmic expansion rate. This method relies on the relative ages of galaxies, 
their peak masses, and passively evolving galaxies to determine the Hubble parameter $H(z)$~\cite{Jimenez_2002}. The Hubble parameter 
is expressed as~\cite{Jimenez_2002,Moresco_2022}

\beqa
H(z)=-\frac{1}{1+z} \frac{\Delta z}{\Delta t}
\eeqa

For CC data, the chi-square statistic is defined as
\beqa
\chi _{\rm CC}^{2}=\Delta H^{T} {\rm Cov}^{-1}_{\rm CC}\Delta H
\eeqa
where $\Delta H =  H^{\rm th}(z) - H^{\rm obs}(z)$. In this work, we use 31 CC data points compiled from various sources~\cite{Moresco_2016,Pal_2024,Moresco_2015,Ratsimbazafy_2017,Zhang_2014}. The covariance matrix $\rm Cov_{CC}$ is computed following the method described 
in Ref.~\cite{Moresco:2020fbm} \footnote{The related code is available at $\rm https://github.com/Ahmadmehrabi/Cosmic\_chronometer\_data$}.

%%%%%%%%%%%%%%%%%%%%%%%%%%%%

\subsection{Type Ia supernova}
Type Ia supernovae (SNIa) are widely utilized as standard candles for probing cosmological models. 
In a flat universe, the luminosity distance is given by
 
\beqa
d_{L}(z) = (1+z)\int^{z}_{0}\frac{dz^{'}}{H(z^{'})/H_0}.
\eeqa

The distance modulus $\mu(z)$ is then defined as

\beqa
\mu(z) = 5{\rm log}_{10}d_{L}(z)+25.
\eeqa
 
For SNIa data, the observed apparent magnitude $m_{\rm obs}$ is related to the observed distance modulus by

\beqa
m_{\rm obs} = \mu_{\rm obs} + M,
\eeqa
where \(M\) denotes the absolute magnitude (in the B-band), which can be calibrated through alternative methods. 
However, in this work, we do not adopt a pre-calibrated value of $M$ (as suggested, e.g., in 
Refs.~\cite{Camarena:2019moy,Camarena:2021jlr,Kang:2019azh,vonMarttens:2025dvv,Lemos:2025qyh,Riess:2019cxk,Chander:2025bml,Camarena:2023rsd}. 
Instead, we use a marginalization technique to handle this nuisance parameter 
when deriving cosmological constraints~\cite{Conley2011SUPERNOVACA,Bouali:2019whr,Gong:2007se,PhysRevD.72.123519,2001A&A...380....6G}. 
The $\chi^{2}$ statistic for this marginalized method applied 
to SNIa data can be expressed as~\cite{Bouali:2019whr,Conley2011SUPERNOVACA}:

\beqa
\chi^{2}_{\rm SNIa}=A-\frac{B^{2}}{C}+{\rm ln}\frac{C}{2\pi},
\eeqa

where 

\beqa
&&A = \Delta \mu^{T}C^{-1}_{\rm SNIa}\Delta \mu \nonumber \\
&&B = \Delta \mu^{T}C^{-1}_{\rm SNIa}I \nonumber \\
&&C = I^{T}C^{-1}_{\rm SNIa}I.
\eeqa
Here, $\Delta \mu = \mu^{\rm th}-\mu^{\rm obs}$, $I$ is the identity matrix, and 
$C_{\rm SNIa}$ is the covariance matrix of the SNIa data, which includes both statistical and systematic uncertainties. 

For our analysis, three SNIa datasets are employed, as detailed below: 
(i) The PantheonPlus sample~\footnote{\url{https://github.com/PantheonPlusSH0ES/DataRelease}}, 
which consists of 1701 light curves from 1550 distinct SNIa, covering a redshift range of $0.001<z<2.26$~\cite{Scolnic:2021amr,Brout:2022vxf}. 
To reduce the impacts of peculiar velocities in nearby galaxies, 
we restrict our analysis to data with $z>0.01$. 
(ii) The five-year SNIa data from the Dark Energy Survey (DES)~\footnote{\url{https://github.com/des-science/DES-SN5YR}}, 
which includes 1635 SNIa in the redshift range $0.1<z<1.3$, 
supplemented by 194 low-redshift SNIa in $0.025<z<0.1$. 
This combined dataset, totaling 1829 SNIa, is commonly known as 
DESY5~\cite{DES:2024jxu,2009ApJ...700..331H,2012ApJS..200...12H,Krisciunas:2017yoe,Foley:2017zdq}. 
(iii) the Union3 sample, which contains 2087 SNIa~\cite{Rubin:2023jdq}. For our calculations, 
we use the binned data~\footnote{\url{https://github.com/CobayaSampler/sn_data}}, 
spanning $0.05<z<2.26$. 

%%%%%%%%%%%%%%%%%%%%%%%%%%%%%%%%%%%%

\section{constraints on the models}

We employ the Markov Chain Monte Carlo (MCMC) method to determine the best-fit values and posterior distributions of 
the parameters for different models: 

(i) $\alpha$CDM model: \{$\Omega_{m}$, $H_0$, $\Omega_{b}h^2$, $r_d$, $\alpha$\};

(ii) $\alpha w$CDM model: \{$\Omega_{m}$, $H_0$, $\Omega_{b}h^2$, $r_d$, $w$, $\alpha$\};

(iii) $\alpha w_0 w_a$CDM model: \{$\Omega_{m}$, $H_0$, $\Omega_{b}h^2$, $r_d$, $w_0$, $w_a$, $\alpha$\};

(iv) $J$CDM model: \{$\Omega_{m}$, $H_0$, $\Omega_{b}h^2$, $r_d$\};

Given the datasets described in the previous section, the total $\chi^{2}$ is given by 

\beqa
\chi^{2}_{\rm total}=\chi^{2}_{\rm BAO} + \chi^{2}_{\rm CMB} + \chi^{2}_{\rm CC} + \chi^{2}_{\rm SNIa},
\eeqa
then the likelihood function is $L\propto e^{-\chi^{2}_{\rm total}/2}$. We perform the MCMC sampling using the 
public code $\mathtt{emcee}$~\cite{emcee}, with uniform priors on the parameters: 
$\Omega_{m}\in (0,1)$, $H_{0}\in (50,90)$ and $\Omega_{b}h^{2} \in (0.0001,0.1)$, $r_d \in (130,170)$, $w \in (-2,2)$, $w_0 \in (-2,2)$, $w_a \in (-2,2)$; 
Constraints on redshift-evolving Hubble constant models with early- and late-time diagnostics
and $\alpha \in (-1,1)$. The MCMC chains are analyzed using $\mathtt{GetDist}$~\cite{getdist}, and the best-fit parameters derived from different datasets 
with their $1\sigma$ uncertainties are summarized in Tabs.~\ref{tab:parameters_total} and~\ref{tab:sep_parameters_total}. 
One-dimensional marginalized probability distributions and two-dimensional confidence contour plots are shown in 
Figs.~\ref{fig:alpha_lambda_cdm} $-$ \ref{fig:sep_jcdm}.

\subsection{Constraints from the combined datasets}

\begin{figure}[!htb]
\includegraphics[width=\linewidth]{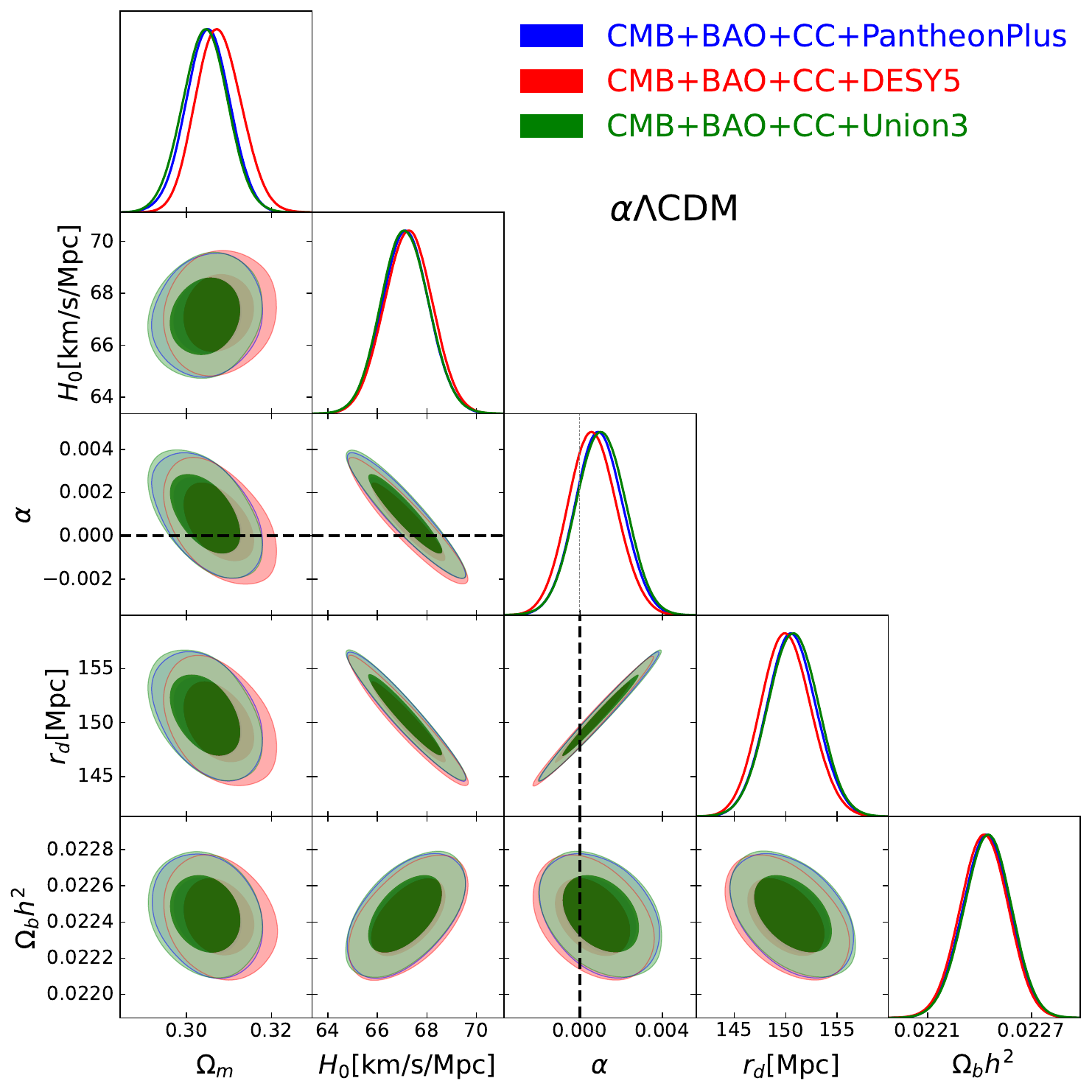}
\caption{Marginalized one-dimensional distributions and two-dimensional confidence contours (68\% and 95\% C.L.) for the parameters of the $\alpha\Lambda$CDM model, using the combined datasets CMB+BAO+CC+PantheonPlus (blue), CMB+BAO+CC+DESY5 (red), and CMB+BAO+CC+Union3 (green). 
The dashed line marks $\alpha=0$.}

\label{fig:alpha_lambda_cdm}
\end{figure}

\begin{figure}[!htb]

%\centering
\includegraphics[width=\linewidth]{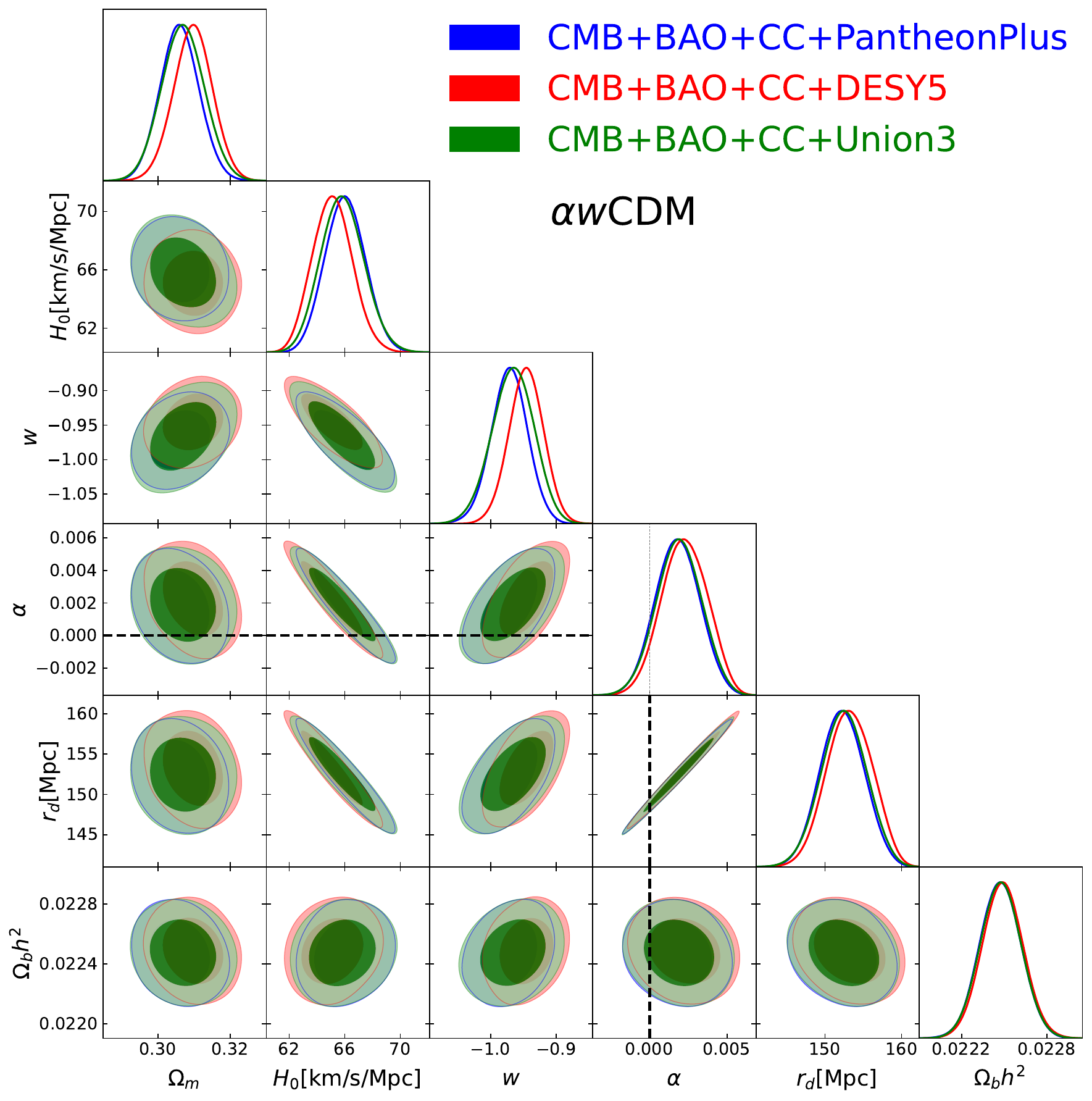}

\caption{Marginalized one-dimensional distributions and two-dimensional confidence contours (68\% and 95\% C.L.) for the parameters of the 
$\alpha w$CDM model, using the combined datasets CMB+BAO+CC+PantheonPlus (blue), CMB+BAO+CC+DESY5 (red), and CMB+BAO+CC+Union3 (green). 
The dashed line marks $\alpha=0$.}
 
\label{fig:alpha_w_cdm}
\end{figure}

\begin{figure}[!htb]

%\centering
\includegraphics[width=\linewidth]{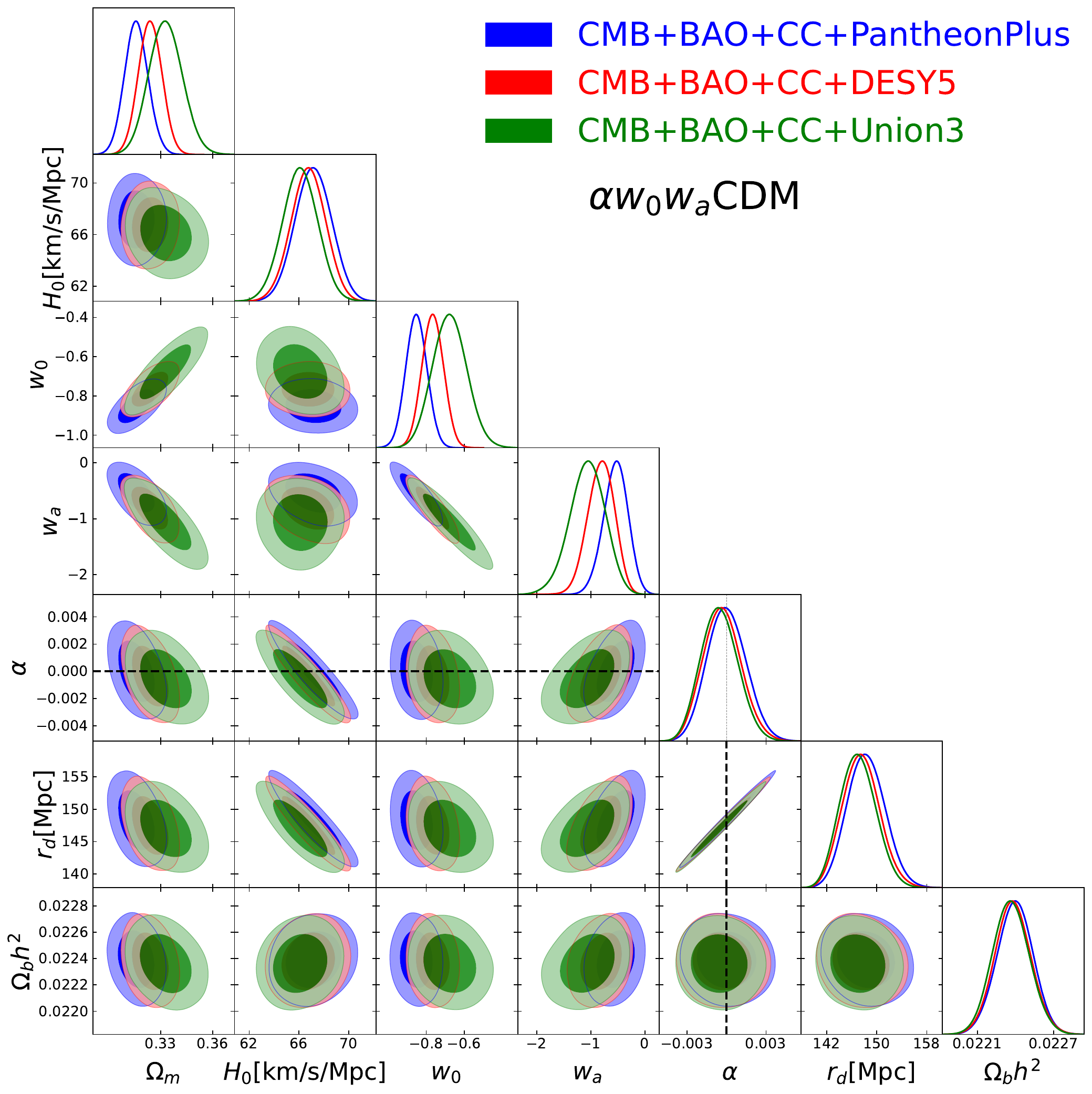}

\caption{Marginalized one-dimensional distributions and two-dimensional confidence contours (68\% and 95\% C.L.) for the parameters of the 
$\alpha w_0 w_a$CDM model, using the combined datasets CMB+BAO+CC+PantheonPlus (blue), CMB+BAO+CC+DESY5 (red), and CMB+BAO+CC+Union3 (green). 
The dashed line marks $\alpha=0$.}

\label{fig:alpha_w0wa_cdm}
\end{figure}

\begin{figure}[!htb]
%\centering

\includegraphics[width=\linewidth]{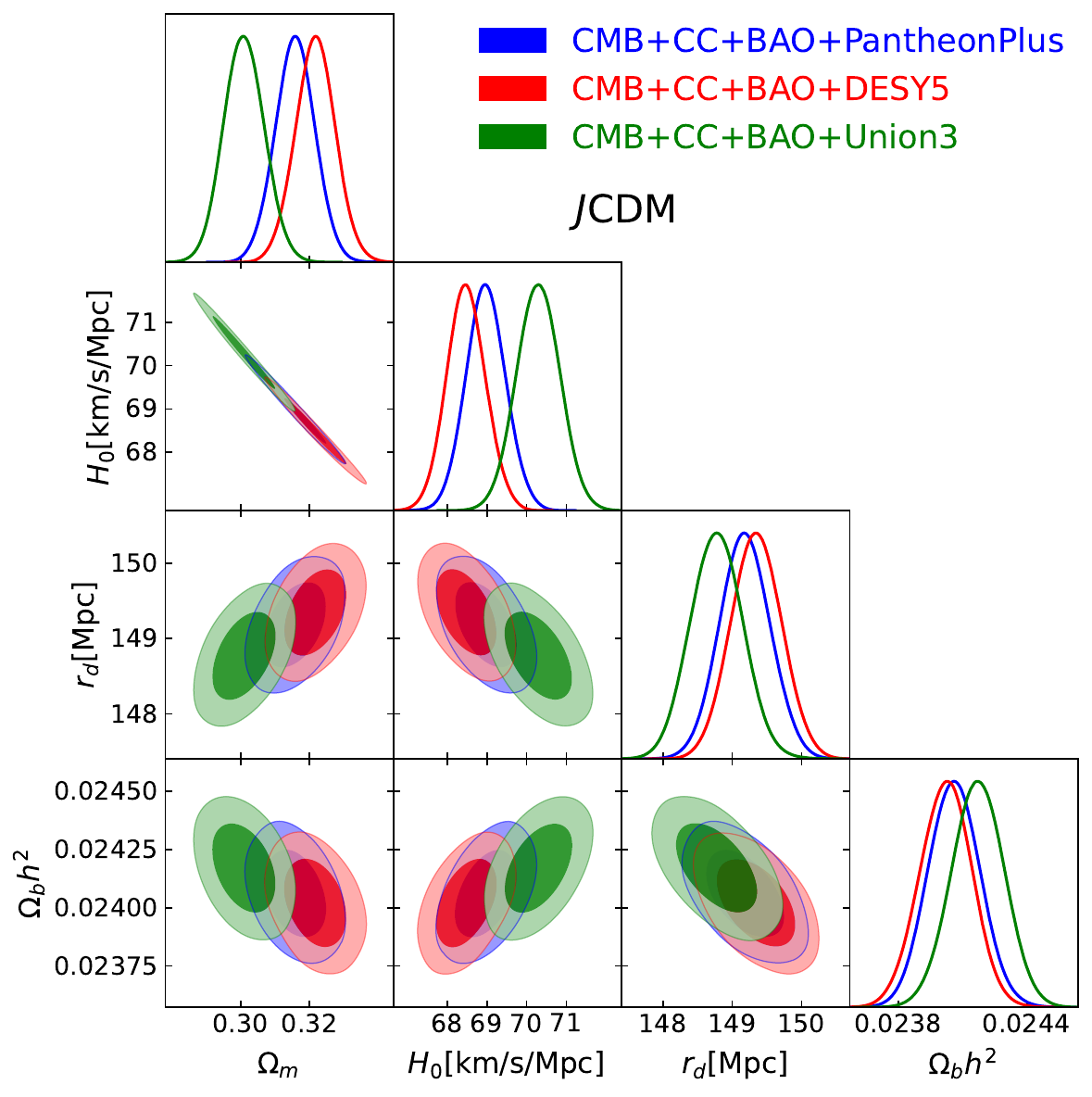}

\caption{Marginalized one-dimensional distributions and two-dimensional confidence contours (68\% and 95\% C.L.) for the parameters of the 
$J$CDM model, using the combined datasets CMB+BAO+CC+PantheonPlus (blue), CMB+BAO+CC+DESY5 (red), and CMB+BAO+CC+Union3 (green).}

\label{fig:jcdm}
\end{figure}

The constraints on the parameters of different models, derived from the total dataset 
CMB+BAO+CC+PantheonPlus/DESY5/Union3, are shown in Tab.~\ref{tab:parameters_total}, and 
one-dimensional marginalized probability distributions and two-dimensional confidence contour plots are shown in 
Figs.~\ref{fig:alpha_lambda_cdm} $-$ \ref{fig:jcdm}.

\begin{figure}[!htb]

%\centering
\includegraphics[width=\linewidth]{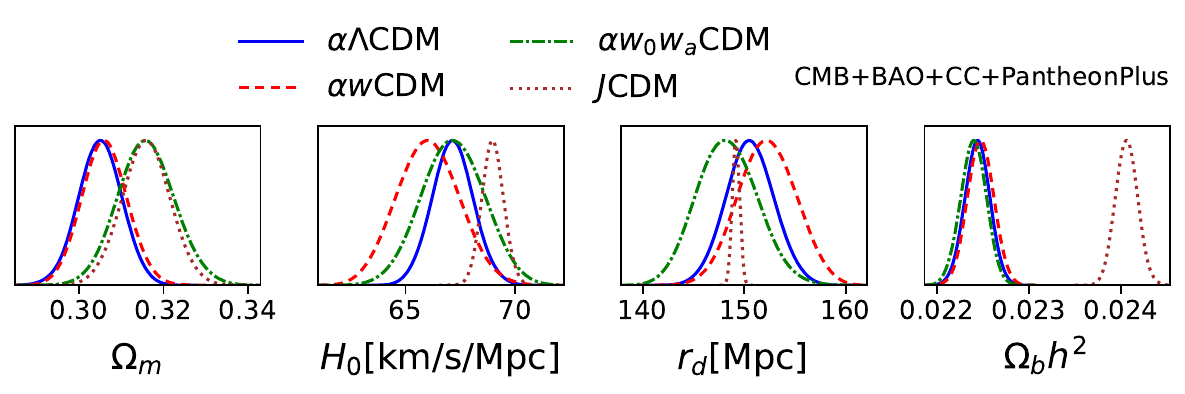}
\includegraphics[width=\linewidth]{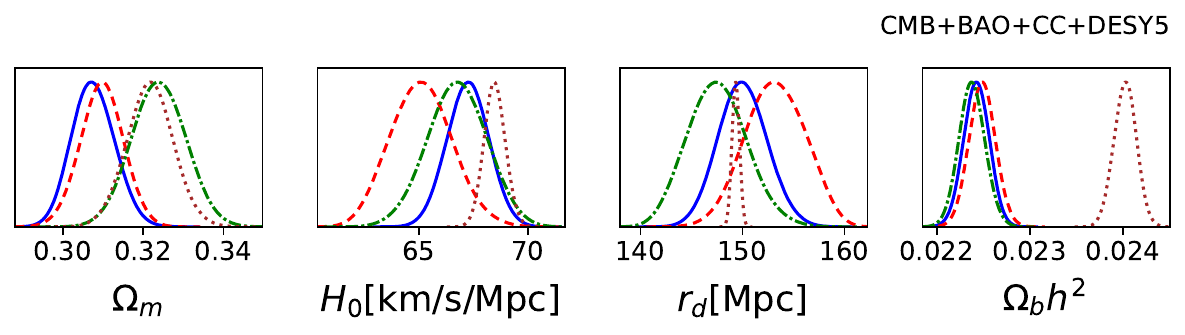}
\includegraphics[width=\linewidth]{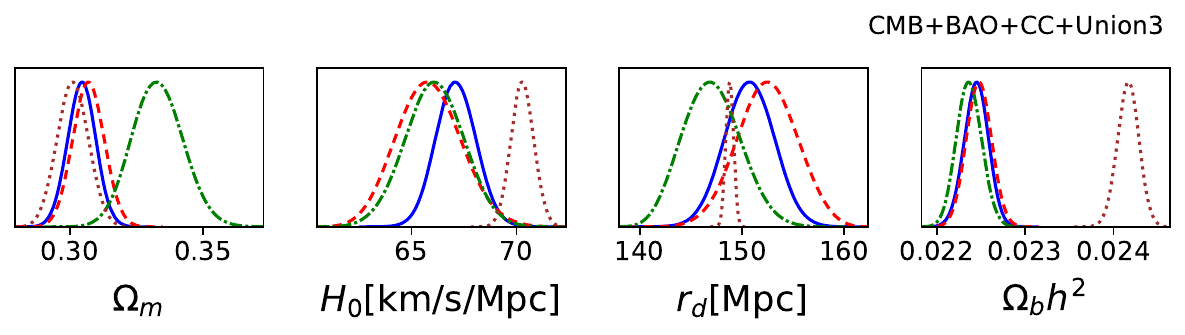}

\caption{Marginalized one-dimensional distributions of $\Omega_m$, $H_0$, $r_d$, and $\Omega_b h^2$ for different models. The three rows correspond to the datasets CMB+BAO+CC+PantheonPlus (top), CMB+BAO+CC+DESY5 (middle), and CMB+BAO+CC+Union3 (bottom).}

\label{fig:compare_parameters}
\end{figure}

The matter density parameter $\Omega_m$ is $\sim 0.30$ 
for the $\alpha \Lambda$CDM and $\alpha w$CDM models across all datasets. For the $J$CDM model, $\Omega_m$ is also $\sim 0.30$ 
for the CMB+BAO+CC+Union3 dataset. For the other models and datasets, $\Omega_m$ is larger than 0.31, especially $\sim 0.33$ in 
the $\alpha w_0 w_a$CDM model for the CMB+BAO+CC+Union3 dataset. One-dimensional marginalized probability distributions of 
$\Omega_m$ for different models for the same dataset are shown in Fig.~\ref{fig:compare_parameters}. From this plot, it can be seen that 
the value of $\Omega_m$ in the $\alpha w_0 w_a$CDM model shows a significant deviation from those in the $\alpha \Lambda$CDM and $\alpha w$CDM models. 

The Hubble constant $H_0$ is around $67~\rm km~s^{-1}~Mpc^{-1}$ in the $\alpha \Lambda$CDM model for all datasets, 
consistent with the value derived from CMB. For the $\alpha w$CDM model, the Hubble constant is smaller for all datasets, 
and the tension with the value derived from the local distance ladder is $4.04\sigma$, $4.55\sigma$, and $3.97\sigma$ for the datasets 
CMB+BAO+CC+PantheonPlus/DESY5/Union3, respectively. For the $\alpha w_0 w_a$CDM model, 
the value of $H_0$ is basically consistent with that derived from CMB for the dataset CMB+BAO+CC+PantheonPlus. 
The values are smaller for the datasets CMB+BAO+CC+DESY5/Union3, with tensions of $3.58\sigma$ and $3.98\sigma$, respectively. 
For the $J$CDM model, the values of $H_0$ are larger than those derived from CMB for all datasets, 
especially $H_0=70.28\pm{0.54}~\rm km~s^{-1}~Mpc^{-1}$ for the CMB+BAO+CC+Union3 dataset. 
The tensions are $3.53\sigma$, $3.97\sigma$, and $2.36\sigma$ for the datasets 
CMB+BAO+CC+PantheonPlus/DESY5/Union3, respectively.  

\begin{figure}[!htb]

%\centering
\includegraphics[width=\linewidth]{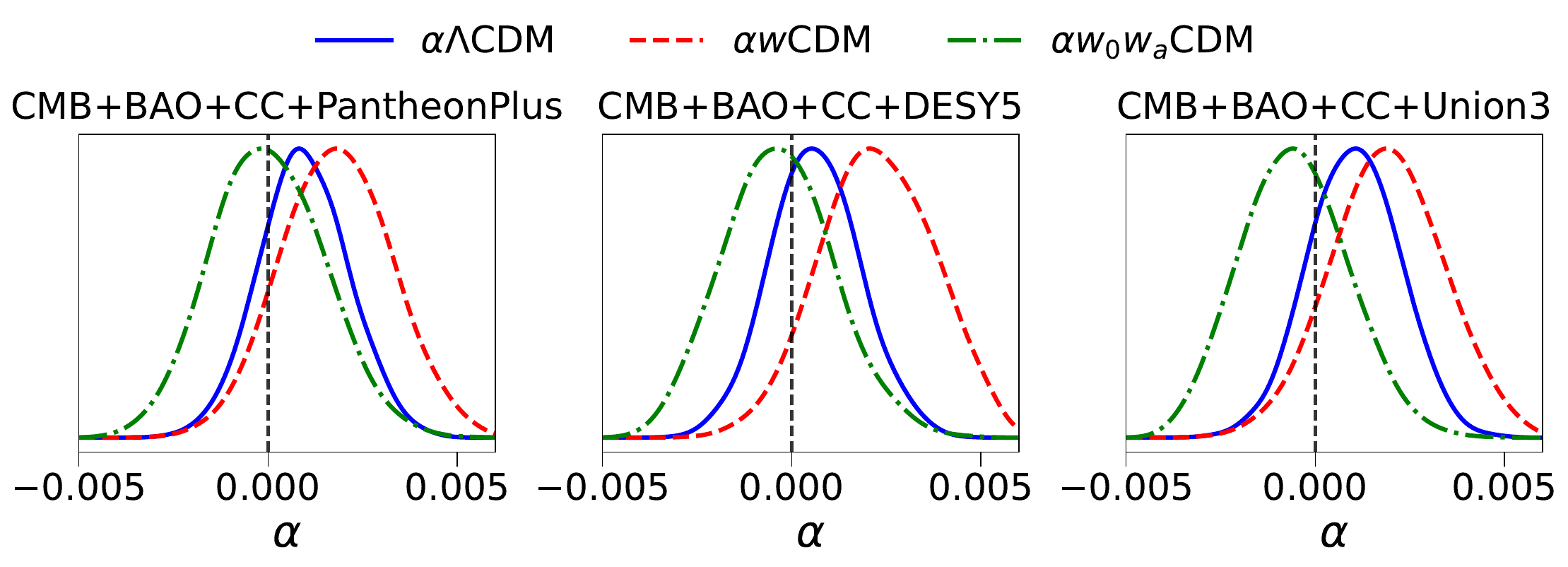}

\caption{Marginalized one-dimensional distributions of the parameter $\alpha$ for different models, using the same dataset (as labeled).}

\label{fig:compare_alpha}
\end{figure}

For the parameter $\alpha$, in the $\alpha$CDM and $\alpha w_0 w_a$CDM models it is consistent with $\alpha=0$ in $1\sigma$, while the 
best-fit value is positive for the $\alpha$CDM model and negative for the $\alpha w_0 w_a$CDM model across all datasets. 
The deviation of $\alpha$ from 0 is significant for the $\alpha w$CDM model, showing a deviation from $\alpha=0$ 
at $1.29\sigma$, $1.64\sigma$, and $1.36\sigma$ for 
the CMB+BAO+CC+PantheonPlus/DESY5/Union3 datasets, respectively, but still consistent with $\alpha=0$ within $2\sigma$. 
One-dimensional marginalized probability distributions of $\alpha$ for different models and the same dataset are shown in Fig.~\ref{fig:compare_alpha}. 
In this plot, the deviation of $\alpha$ from 0 is clearly visible for the $\alpha w$CDM model for all datasets. 
For the dynamic dark energy model $\alpha w_0 w_a$CDM, the evidence is still significant, with $2.4\sigma$, $3.8\sigma$, and $3.5\sigma$ 
for the datasets CMB+BAO+CC+PantheonPlus/DESY5/Union3, respectively. These results are basically consistent with those, e.g., in~\cite{DESI:2025zgx}. 
This result is also attributed to the $\alpha$ value, which is consistent with $\alpha=0$ for all datasets.

The baryon density parameter $\Omega_b h^2$ is consistent across the $\alpha\Lambda$CDM, $\alpha w$CDM, and $\alpha w_0 w_a$CDM models for 
all datasets. These values are also consistent with the Planck 2018 results~\cite{2018Planck}. However, the $J$CDM model prefers a larger value, 
$\Omega_b h^2 \sim 0.24$, as can be clearly seen in Fig.~\ref{fig:compare_parameters}. 

For the acoustic horizon $r_d$, we have treated it as a free parameter. Planck 2018 derived a value of $r_d = 147.09\pm{0.26}\rm Mpc$ 
for the standard $\Lambda$CDM model. Among the models investigated here, the $\alpha w_0 w_a$CDM model shows a result similar to 
that of $\Lambda$CDM model, while larger values are found for the other models. On the other hand, since the errors on $r_d$ for 
the $\alpha\Lambda$CDM, $\alpha w$CDM, and $\alpha w_0 w_a$CDM models are larger, 
their $r_d$ values are essentially consistent with that of $\Lambda$CDM within $2\sigma$. 
For $J$CDM, there are tensions with $\Lambda$CDM model, at $4.3\sigma$, $4.8\sigma$, and $3.5\sigma$ for the datasets 
CMB+BAO+CC+PantheonPlus/DESY5/Union3, respectively.

%****++++++++++++++++++++++++ Table I***************************************

\begin{table*}[htb]
%\footnotesize
\caption{Parameter constraints with 1$\sigma$ uncertainty for different models. From top to bottom, the rows present results for 
the combined datasets CMB+BAO+CC+PantheonPlus, CMB+BAO+CC+DESY5, and CMB+BAO+CC+Union3, respectively.}
\label{tab:parameters_total}
\centering
\scriptsize 
\begin{ruledtabular} 
\begin{tabular}{lccccccccc}

Model & $\Omega_{m}$ & $H_0[\rm km~s^{-1}~Mpc^{-1}]$ &$\alpha$& $w~{\rm or}~w_0$ & $w_a$ & $\Omega_{b}h^{2}$ & $\Omega_{m}h^{2}$&$r_d$[Mpc] \\
\hline
\multirow{3}{*}{\(\alpha \Lambda \rm CDM\)}
   & $0.3050\pm{0.0051}$   & $67.14\pm{0.94}$  & $0.0009\pm{0.0011}$ & ... &...&$0.02244\pm{0.00014}$ & $0.1375\pm{0.0047}$ &$150.6\pm{2.4}$ \\
   & $0.3076\pm{0.0053}$   & $67.25\pm{0.96}$  & $0.0006\pm{0.0012}$ & ... &...&$0.02243\pm{0.00014}$ & $0.1391\pm{0.0048}$ &$150.0\pm{2.4}$ \\   
   & $0.3044\pm{0.0053}$   & $67.12\pm{0.95}$  & $0.0010\pm{0.0012}$ & ... &...&$0.02245\pm{0.00014}$ & $0.1372\pm{0.0049}$ &$150.7\pm{2.4}$ \\  
\hline
\multirow{3}{*}{\(\alpha w \rm CDM\)}
   & $0.3060\pm{0.0053}$   & $66.00\pm{1.40}$  & $0.0018\pm{0.0014}$ & $-0.971\pm{0.028}$ &...&$0.02247\pm{0.00014}$ & $0.1335\pm{0.0060}$ &$152.3\pm{2.8}$ \\
   & $0.3097\pm{0.0053}$   & $65.10\pm{1.40}$  & $0.0023\pm{0.0014}$ & $-0.946\pm{0.026}$ &...&$0.02249\pm{0.00014}$ & $0.1313\pm{0.0060}$ &$153.2\pm{2.9}$ \\   
   & $0.3070\pm{0.0058}$   & $65.80\pm{1.50}$  & $0.0019\pm{0.0014}$ & $-0.966\pm{0.032}$&...&$0.02247\pm{0.00014}$  & $0.1330\pm{0.0062}$ &$152.5\pm{2.9}$ \\ 
\hline
\multirow{3}{*}{\(\alpha w_0 w_a \rm CDM\)}
   & $0.3159\pm{0.0067}$   & $67.20\pm{1.40}$  & $-0.0001\pm{0.0014}$ & $-0.852\pm{0.055}$ &$-0.54^{+0.24}_{-0.21}$&$0.02239\pm{0.00014}$ & $0.1425\pm{0.0068}$ &$148.3\pm{3.0}$ \\
   & $0.3239\pm{0.0067}$   & $66.80\pm{1.40}$  & $-0.0003\pm{0.0014}$ &$-0.766\pm{0.056}$&$-0.81^{+0.26}_{-0.23}$&$0.02238\pm{0.00014}$   & $0.1444\pm{0.0068}$&$147.5\pm{2.9}$ \\   
   & $0.3329\pm{0.0095}$   & $66.10\pm{1.40}$  & $-0.0005\pm{0.0014}$ &$-0.677\pm{0.089}$&$-1.06\pm{0.32}$&$0.02236\pm{0.00014}$          & $0.1455\pm{0.0067}$&$147.0\pm{2.8}$ \\ 
\hline
\multirow{3}{*}{\(J \rm CDM\)}
   & $0.3158\pm{0.0057}$ & $68.97\pm{0.50}$ &...&...&...&$0.02406\pm{0.00012}$ & $0.1502\pm{0.0007}$ &$149.2\pm{0.4}$\\
   & $0.3217\pm{0.0057}$ & $68.48\pm{0.49}$ &...&...&...&$0.02402\pm{0.00012}$ & $0.1508\pm{0.0007}$ &$149.4\pm{0.4}$ \\
   & $0.3010\pm{0.0058}$ & $70.28\pm{0.54}$ &...&...&...&$0.02417\pm{0.00012}$ & $0.1487\pm{0.0007}$ &$148.8\pm{0.4}$ \\
   
\hline
\multirow{3}{*}{\(\Lambda \rm CDM\)}
   & $0.3069\pm{0.0045}$ & $67.86\pm{0.35}$ &...&...&...&$0.02248\pm{0.00013}$ & $0.1413\pm{0.0007}$ &$148.7\pm{0.4}$\\
   & $0.3089\pm{0.0046}$ & $67.71\pm{0.35}$ &...&...&...&$0.02246\pm{0.00013}$ & $0.1416\pm{0.0007}$ &$148.7\pm{0.4}$ \\
   & $0.3066\pm{0.0048}$ & $67.89\pm{0.36}$ &...&...&...&$0.02248\pm{0.00012}$ & $0.1413\pm{0.0008}$ &$148.6\pm{0.4}$ \\

\end{tabular}
\end{ruledtabular}
\end{table*}

%%++++++++++++++++++++++++++ figure1 alpha Lambda CDM ++++++++++++++++++++++++++++++++++++++
%%++++++++++++++++++++++++++ figure1 ++++++++++++++++++++++++++++++++++++++

%%++++++++++++++++++++++++++ figure2  alpha w CDM ++++++++++++++++++++++++++++++++++++++

%+++++++++++++++++++++++++++++++ figure2 ++++++++++++++++++++++++++++++++++++++++%

%%++++++++++++++++++++++++++ figure3  alpha w0wa CDM ++++++++++++++++++++++++++++++++++++++
%%++++++++++++++++++++++++++ figure3 ++++++++++++++++++++++++++++++++++++++

%%++++++++++++++++++++++++++ figure4 ++++++++++++++++++++++++++++++++++++++
%%++++++++++++++++++++++++++ figure4 ++++++++++++++++++++++++++++++++++++++

%%++++++++++++++++++++++++++ figure 5 compare paramters ++++++++++++++++++++++++++++++++++++++
%%++++++++++++++++++++++++++ figure5 ++++++++++++++++++++++++++++++++++++++

%%%++++++++++++++++++++++++++++++++++++++++++++++++++++++++++++++++++++++++++

\subsection{Constraints from early- and late-time datasets}

\begin{figure}[!htb]
\includegraphics[width=\linewidth]{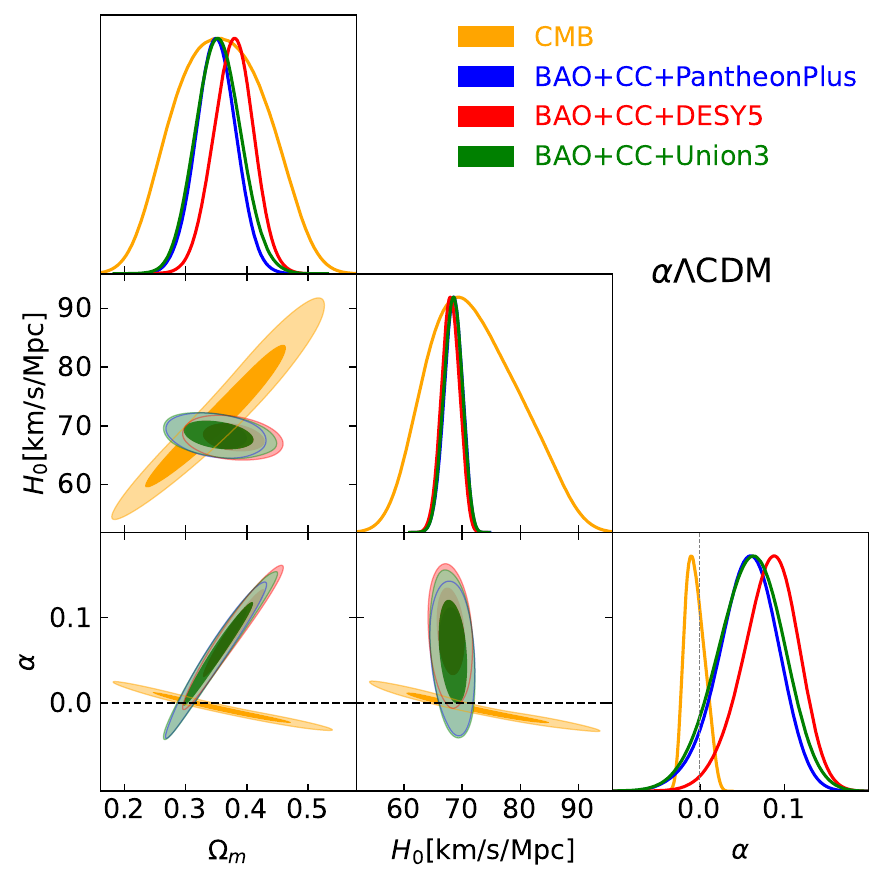}

\caption{Marginalized one-dimensional distributions and two-dimensional contours (68\% and 95\% C.L.) for the $\alpha\Lambda$CDM model, 
using early-time CMB data (orange) and three late-time datasets: BAO+CC+PantheonPlus (blue), BAO+CC+DESY5 (red), and BAO+CC+Union3 (green). 
The dashed line indicates $\alpha=0$.}

\label{fig:sep_alpha_lambda_cdm}
\end{figure}

\begin{figure}[!htb]
\includegraphics[width=\linewidth]{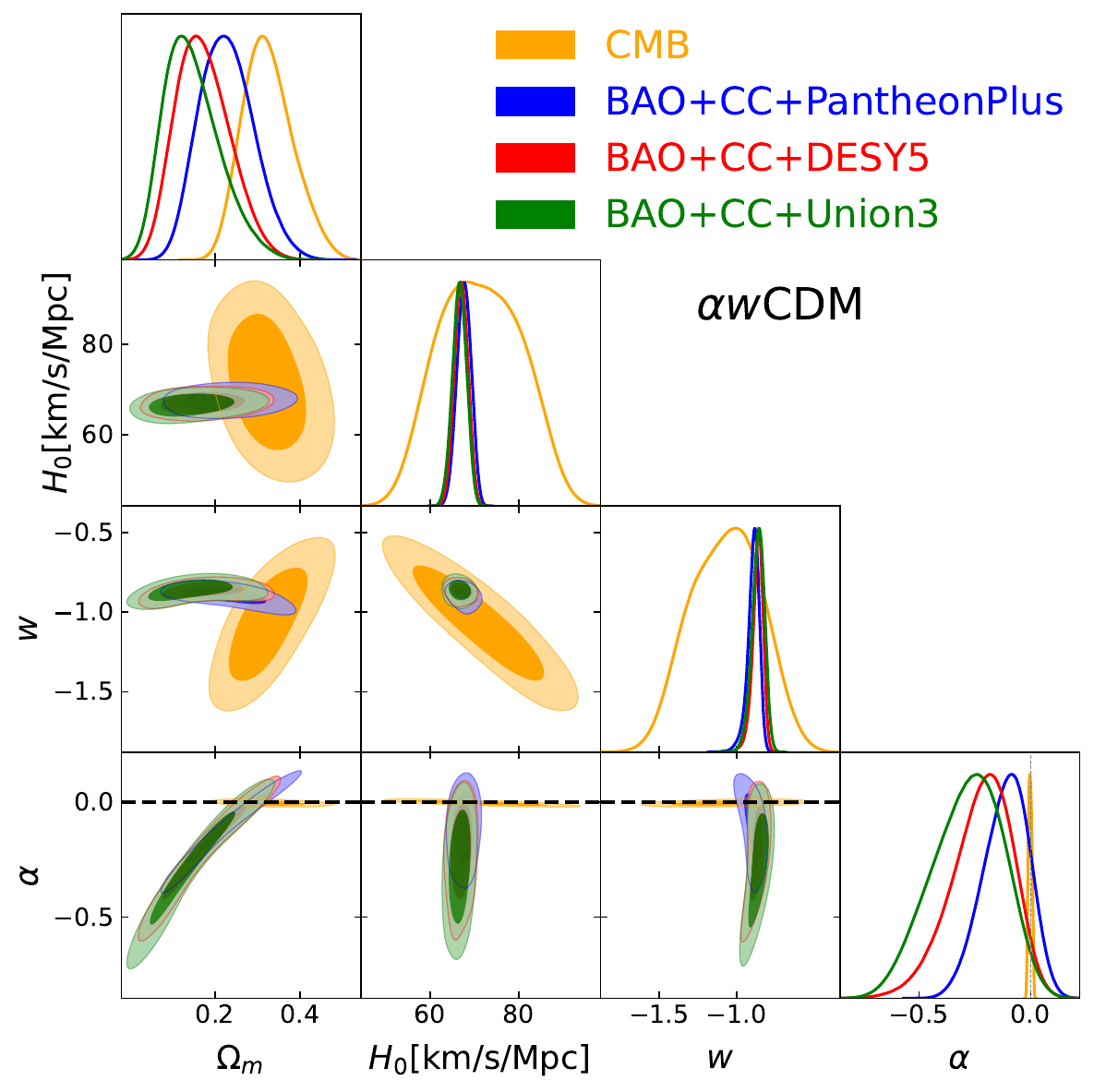}

\caption{Marginalized one-dimensional distributions and two-dimensional contours (68\% and 95\% C.L.) for the $\alpha w$CDM model, 
using early-time CMB data (orange) and three late-time datasets: BAO+CC+PantheonPlus (blue), BAO+CC+DESY5 (red), and BAO+CC+Union3 (green). 
The dashed line indicates $\alpha=0$.}

\label{fig:sep_alpha_w_cdm}
\end{figure}

\begin{figure}[!htb]
\includegraphics[width=\linewidth]{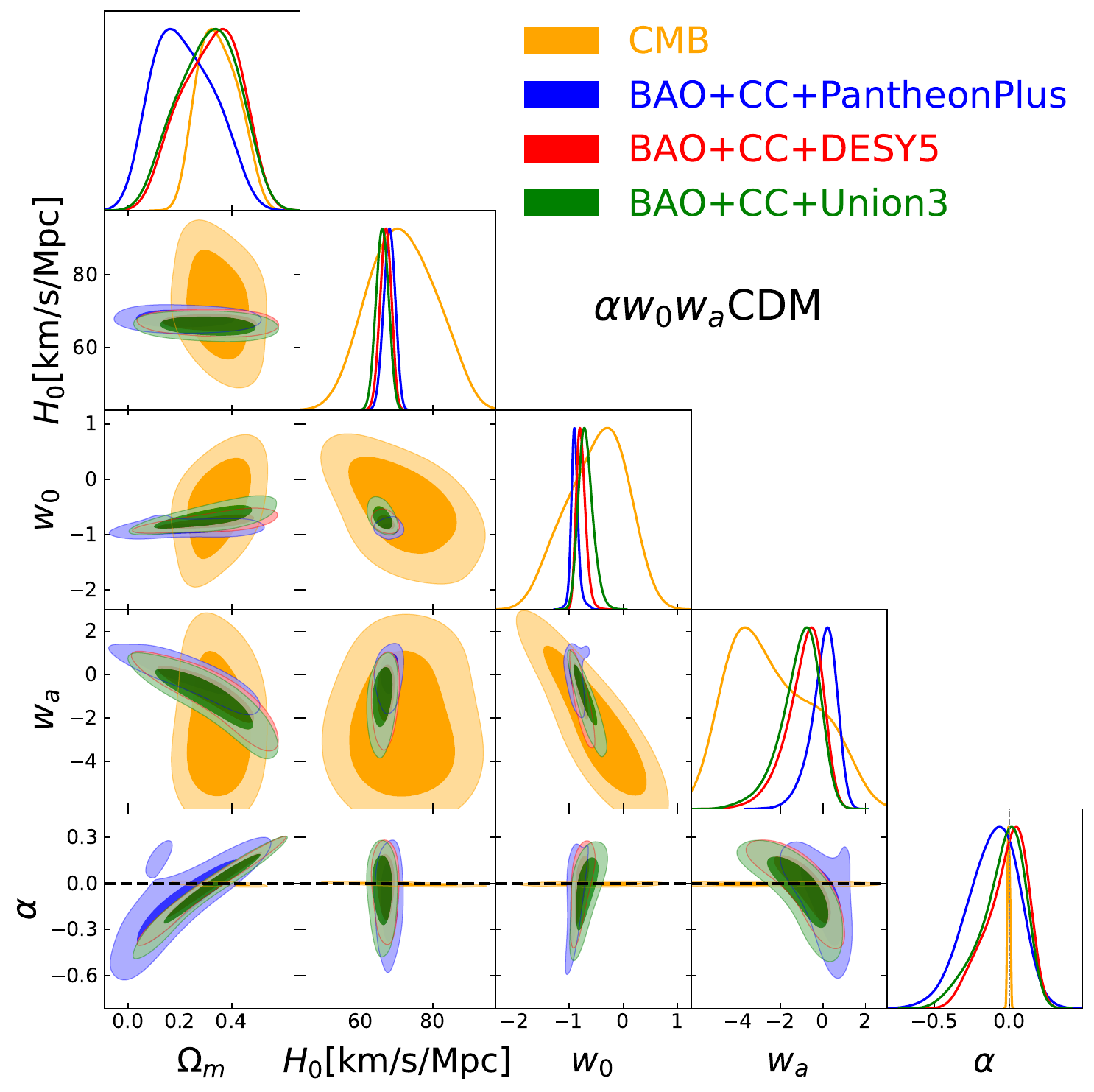}

\caption{Marginalized one-dimensional distributions and two-dimensional contours (68\% and 95\% C.L.) for the $\alpha w_0 w_a$CDM model, 
using early-time CMB data (orange) and three late-time datasets: BAO+CC+PantheonPlus (blue), BAO+CC+DESY5 (red), and BAO+CC+Union3 (green). 
The dashed line indicates $\alpha=0$.}

\label{fig:sep_w0wa_cdm}
\end{figure}

\begin{figure}[!htb]
\includegraphics[width=\linewidth]{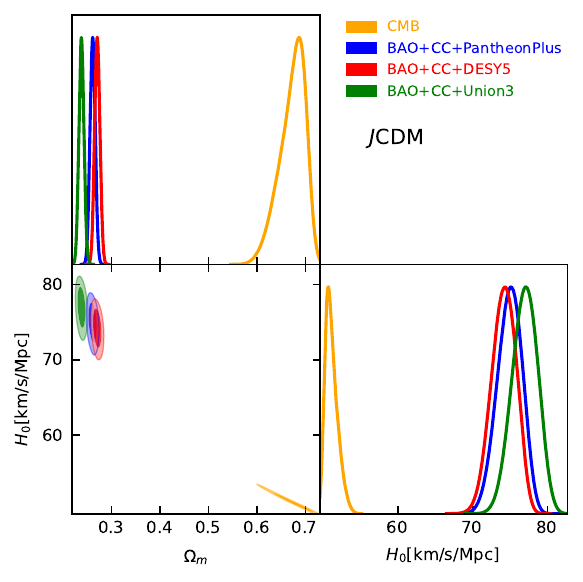}

\caption{Marginalized one-dimensional distributions and two-dimensional contours (68\% and 95\% C.L.) for the $J$CDM model, 
using early-time CMB data (orange) and three late-time datasets: BAO+CC+PantheonPlus (blue), BAO+CC+DESY5 (red), and BAO+CC+Union3 (green). 
}

\label{fig:sep_jcdm}
\end{figure}

Although all datasets give robust constraints on the parameters, it is important to investigate the impact 
of different epoch data on the model constraints, especially for the datasets used in this work which can be 
divided into two epochs: one is early-time CMB (data) and the other is late-time data (BAO+CC+SNIa). 
The constraints on the parameters of different models, derived from the early-time dataset (CMB, first row for each model) 
and late-time datasets 
(BAO+CC+PantheonPlus/DESY5/Union3, subsequent rows), are shown in Tab.~\ref{tab:sep_parameters_total}, and 
one-dimensional marginalized probability distributions and two-dimensional confidence contour plots are shown in 
Figs.~\ref{fig:sep_alpha_lambda_cdm} $-$ \ref{fig:sep_jcdm}. 

For the models $\alpha \Lambda$CDM, $\alpha w$CDM, and $\alpha w_0 w_a$CDM, the central values of $H_0$ derived from 
CMB, the early-time data, are $\sim 71-72~ \rm km~s^{-1}~Mpc^{-1}$ but with large uncertainties ($\sigma \sim 9$). 
For the constraints from the datasets BAO+CC+SNIa, corresponding to the late-time data, the central values of $H_0$ 
are $\sim 66-68~ \rm km~s^{-1}~Mpc^{-1}$ with smaller uncertainties ($\sigma \sim 1.5$). The central values shift 
downward by $\sim 4-5~ \rm km~s^{-1}~Mpc^{-1}$, but due to the large CMB errors, the differences are statistically insignificant. 
These features can also be found in Figs.~\ref{fig:sep_alpha_lambda_cdm},~\ref{fig:sep_alpha_w_cdm}, and~\ref{fig:sep_w0wa_cdm}. 
The values of $\alpha$ are consistent with zero within $2\sigma$ for all the models and datasets. 
From Figs.~\ref{fig:sep_alpha_lambda_cdm},~\ref{fig:sep_alpha_w_cdm}, and~\ref{fig:sep_w0wa_cdm}, it can be seen that 
there is a strong correlation among parameters such as $H_0$, $\alpha$, $w$, and $w_0$, 
which is an important factor contributing to the fitting results. Note that for $\alpha w_0 w_a$CDM model, the late-time values of $H_0$ are similar to the combined result (all datasets, Tab.~\ref{tab:parameters_total}), and this model can not alleviate the Hubble tension. 
 
For the $\alpha \Lambda$CDM model, the central value of $\alpha$ from the CMB-only fit is close to zero 
($\alpha =-0.0057$, Tab.~\ref{tab:sep_parameters_total}), whereas the posterior values of $\Omega_m$ and $H_0$ are larger than those 
of the Planck $\Lambda$CDM model. This displacement arises because the CMB distance priors constrain the 
combinations $l_a \sim D_A(z_{\star})/r_s(z_\star)$ (Eq.~(\ref{eq:cmb_la})) and 
$R\sim \sqrt{\Omega_mH^2_0}D_A(z_\star)$ (Eq.~(\ref{eq:cmb_R})), rather than $H_0$ and $\Omega_m$ individually. 
For $\alpha <0$, the factor $(1+z_\star)^{-2\alpha}$ enhances the early-time expansion rate and reduces $r_s(z_\star)$. 
To keep $l_a$ at the Planck value, $D_A(z_\star)$ must also decrease, which is achieved by increasing $\Omega_m$. 
The shift parameter $R$ then requires a compensating increase in $H_0$. The joint solution of the two priors therefore moves 
the posterior centre to $\alpha = -0.0057$, $\Omega_m = 0.358$, and $H_0 = 72.1$, 
which in turn gives $\Omega_mh^2 = 0.196$. Moreover, this $H_0-\Omega_m$ degeneracy is clearly visible 
in Fig.~\ref{fig:sep_alpha_lambda_cdm}: the marginalized two-dimensional contours are strongly elongated 
along a direction that simultaneously increases $H_0$ and $\Omega_m$, and the large central values quoted 
in Tab.~\ref{tab:sep_parameters_total} are simply the centres of these elongated contours rather than being independently determined. 
When BAO+CC+SNIa data are added, the low-redshift expansion rate and distances break this degeneracy and pull the fit back to 
values consistent with Planck. The CMB-only row should therefore be read as the degeneracy direction of 
the distance priors under $\alpha\Lambda$CDM, rather than as evidence for a deviation from $\Lambda$CDM.

For the $J$CDM model, it shows a dramatic discrepancy between the early- and late-time constraints. The CMB-only fit yields 
$H_0 = 51.03^{+0.57}_{-0.99}~ \rm km~s^{-1}~Mpc^{-1}$  and $\Omega_m =0.674^{+0.032}_{-0.019}$, both of which are in 
tension ($\sim 10\sigma$) with Planck $\Lambda$CDM values. In contrast, the late-time fit gives $H_0 = 74.9^{+1.8}_{-1.6}~ \rm km~s^{-1}~Mpc^{-1}$ (BAO+CC+PantheonPlus),  
$H_0 = 74.2^{+1.8}_{-1.6}~ \rm km~s^{-1}~Mpc^{-1}$ (BAO+CC+DESY5), $H_0 = 77.0^{+1.8}_{-1.6}~ \rm km~s^{-1}~Mpc^{-1}$ (BAO+CC+Union3), 
which are close to the SH0ES value and reduce the tension to $0.93\sigma$, $0.58\sigma$, 
and $1.98\sigma$, respectively. Moreover, the constraints on $\Omega_mh^2$ are $0.1466^{+0.0068}_{-0.0061}$ (BAO+CC+PantheonPlus), 
$0.1488\pm{0.0063}$ (BAO+CC+DESY5), and $0.1409\pm{0.0061}$ (BAO+CC+Union3), respectivey, and these results are in excellent agreement with Planck 
results $\Omega_mh^2 = 0.1424\pm{0.00087}$ within $1\sigma$. However, when CMB is included 
(Tab.~\ref{tab:parameters_total}, CMB+BAO+CC+Union3), $H_0$ is decreased to $\sim 70.3$, and $\Omega_b h^{2}$ is derived as $0.02417\pm{0.00012}$, deviating from the Planck value by more than $5\sigma$. These features can also be found in Fig.~\ref{fig:sep_jcdm}. 
This demonstrates that $J$CDM alleviates the late-time tension at the cost of breaking early-universe consistency, notably in 
$\Omega_bh^2$ (Tab.~\ref{tab:parameters_total}), while preserving the Planck BAO constraint $\Omega_mh^2$ (Tab.~\ref{tab:sep_parameters_total}). 
This is the reason it is strongly rejected by AIC/BIC (see the following subsection). This situation is consistent with the results presented in Ref.~\cite{Wang:2026rkx}.

To quantify the redistribution of the tension, we take the Planck $\Lambda$CDM values $H_0 = 67.4~\rm km~s^{-1}~Mpc^{-1}$, 
$\Omega_{b}h^2=0.02237$, and $\Omega_{m}h^2=0.1424$, and the SH0ES value $H_0 = 73.04~\rm km~s^{-1}~Mpc^{-1}$. The original tension is 
$(73.04-67.4)/67.4=8.4\%$. For $\alpha \Lambda$CDM, the combined fits give $H_0 = 67.14, 67.25, 67.12$, i.e., -0.4\% to -0.2\% relative to Planck, 
while $\Omega_{b}h^2$ is within 0.4\% and $\Omega_{m}h^2$ is lower by 2.3\%$-$3.7\%. Thus the $H_0$ tension is essentially unchanged. 
For $\alpha w$CDM, $H_0 = 66.00, 65.10, 65.80$, i.e., 2.1\%$-$3.4\% below Planck, while $\Omega_{m}h^2$ is lower by 6.3\%$-$7.8\%; 
the tension is not alleviated but worsened in $H_0$. For $\alpha w_0 w_a$CDM, $H_0 = 67.20, 66.80, 66.10$, i.e., 0.3\%$-$1.9\% below Planck, $\Omega_bh^2$ is 
within 0.1\%, and $\Omega_mh^2$ is shifted upward by 0.1\%$-$2.2\%;  the marginal AIC preference does not increase $H_0$. 
$J$CDM shows a significant redistribution: $H_0$ is 2.3\%$-$4.3\% above Planck 
but still 3.8\%$-$6.2\% below SH0ES, while $\Omega_bh^2$ and $\Omega_mh^2$ are shifted upward by 7.4\%$-$8.0\% and 4.4\%$-$5.9\%, respectively. 
Therefore, in $J$CDM the $H_0$ tension is partly transferred to $\Omega_bh^2$ and $\Omega_mh^2$.

%****++++++++++++++++++++++++ Table II  ***************************************

\begin{table*}[htb]
%\footnotesize
\caption{Parameter constraints with 1$\sigma$ uncertainty for different models obtained using early- and late-time data individually. From top to bottom, the four rows list the results for the CMB, BAO+CC+PantheonPlus, BAO+CC+DESY5, and BAO+CC+Union3 datasets, respectively.}
\label{tab:sep_parameters_total}
\centering
\begin{ruledtabular} 
\begin{tabular}{lcccccccc}

Model & $\Omega_{m}$ & $H_0[\rm km~s^{-1}~Mpc^{-1}]$ &$\alpha$& $w~{\rm or}~w_0$ & $w_a$ & $\Omega_{b}h^2$ & $\Omega_{m}h^2$\\
\hline
\multirow{3}{*}{\(\alpha \Lambda \rm CDM\)}
   & $0.358\pm{0.071}$   & $72.1^{+6.7}_{-9.1}$  & $-0.0057^{+0.0097}_{-0.013}$ & ... &...& $0.02235\pm{0.00019}$ &$0.196^{+0.060}_{-0.097}$ \\
   & $0.350\pm{0.032}$   & $68.4\pm{1.5}$  & $0.057^{+0.037}_{-0.033}$ & ... &...& $0.0180\pm{0.0030}$&$0.164\pm{0.015}$ \\   
   & $0.378\pm{0.032}$   & $68.0\pm{1.5}$  & $0.083^{+0.036}_{-0.030}$ & ... &...& $0.0160\pm{0.0025}$ &$0.175\pm{0.015}$ \\
   & $0.354\pm{0.037}$   & $68.4\pm{1.6}$  & $0.060^{+0.041}_{-0.037}$ & ... &...& $0.0178\pm{0.0033}$ &$0.165\pm{0.016}$ \\  
\hline
\multirow{3}{*}{\(\alpha w \rm CDM\)}
   & $0.324^{+0.051}_{-0.067}$   & $71\pm{9}$  & $-0.0028\pm{0.0075}$ & $-1.07\pm{0.23}$ &...                & $0.02235\pm{0.00019}$ &$0.165^{+0.035}_{-0.053}$ \\
   & $0.227^{+0.058}_{-0.070}$   & $67.70\pm{1.60}$  & $-0.11^{+0.11}_{-0.09}$ & $-0.89^{+0.04}_{-0.03}$ &...& $0.049^{+0.016}_{-0.032}$ &$0.10\pm{0.03}$ \\   
   & $0.172^{+0.055}_{-0.072}$   & $67.00\pm{1.50}$  & $-0.22^{+0.15}_{-0.11}$ & $-0.86^{+0.04}_{-0.03}$ &...& $0.100^{+0.042}_{-0.090}$ &$0.08\pm{0.03}$ \\
   & $0.145^{+0.051}_{-0.076}$   & $66.70\pm{1.60}$  & $-0.28^{+0.18}_{-0.14}$ & $-0.86^{+0.04}_{-0.04}$ &...& $0.162^{+0.072}_{-0.17}$ &$0.07^{+0.02}_{-0.04}$ \\ 
\hline
\multirow{3}{*}{\(\alpha w_0 w_a \rm CDM\)}
   & $0.348^{+0.074}_{-0.089}$& $72\pm{9}$  & $-0.0046^{+0.0079}_{-0.0092}$ & $-0.49^{+0.64}_{-0.50}$ &$-2.4^{+1.7}_{-2.6}$ & $0.02238^{+0.00019}_{-0.00017}$ &$0.178^{+0.042}_{-0.063}$ \\
   & $0.221^{+0.108}_{-0.141}$& $67.9\pm{1.6}$  & $-0.11^{+0.19}_{-0.15}$ & $-0.89^{+0.05}_{-0.07}$ &$0.03^{+0.71}_{-0.46}$ & $0.061^{+0.040}_{-0.059}$ &$0.0996^{+0.050}_{-0.064}$\\   
   & $0.322^{+0.131}_{-0.109}$& $66.8\pm{1.5}$  & $-0.02^{+0.17}_{-0.10}$ & $-0.78^{+0.07}_{-0.01}$ &$-0.89^{+1.0}_{-0.61}$ & $0.039^{+0.041}_{-0.059}$ &$0.143^{+0.059}_{-0.048}$\\
   & $0.312^{+0.132}_{-0.114}$& $65.9\pm{1.7}$  & $-0.05^{+0.18}_{-0.10}$ & $-0.69^{+0.11}_{-0.15}$ &$-1.09^{+1.1}_{-0.67}$    & $0.046^{+0.031}_{-0.061}$ &$0.134^{+0.056}_{-0.048}$\\ 
\hline
\multirow{3}{*}{\(J \rm CDM\)}
   & $0.674^{+0.032}_{-0.019}$ & $51.03^{+0.57}_{-0.99}$ &...&...&...&$0.02244\pm{0.00016}$ &$0.1753^{+0.0017}_{-0.0012}$\\
   & $0.2609\pm{0.0053}$ & $74.9^{+1.8}_{-1.6}$ &...&...&...         &$0.0224\pm{0.0020}$   &$0.1466^{+0.0068}_{-0.0061}$\\
   & $0.2705\pm{0.0053}$ & $74.2^{+1.8}_{-1.6}$ &...&...&...         &$0.0221\pm{0.0019}$   &$0.1488\pm{0.0063}$\\
   & $0.2377\pm{0.0051}$ & $77.0^{+1.8}_{-1.6}$ &...&...&...         &$0.0234\pm{0.0021}$   &$0.1409\pm{0.0061}$\\

\hline
\multirow{3}{*}{\(\Lambda \rm CDM\)}
   & $0.3156\pm{0.0093}$ & $67.23\pm{0.67}$    &...&...&...&$0.02237\pm{0.00018}$ &$0.1426\pm{0.0014}$\\
   & $0.3009\pm{0.0060}$ & $68.9\pm{1.5}$      &...&...&...&$0.0390\pm{0.0037}$   &$0.1428\pm{0.0065}$\\
   & $0.3048\pm{0.0061}$ & $68.7\pm{1.5}$      &...&...&...&$0.0395\pm{0.0036}$   &$0.1438\pm{0.0064}$\\
   & $0.3005\pm{0.0062}$ & $68.9\pm{1.5}$      &...&...&...&$0.0390\pm{0.0036}$   &$0.1427\pm{0.0064}$\\

\end{tabular}
\end{ruledtabular}
\end{table*}

%%%%%%%%%%%%%%%  figures, separate datasets  begin %%%%%%%%%%%%%%%%%%%%%%%%%%%%%

%%%%%%%%%%%%%%%%%%%%%%%%% figure separate datasets END **************************************

\subsection{Model comparison using AIC and BIC}

We conduct AIC and BIC analyses for model comparison. The AIC and BIC are defined as~\cite{Akaike1974A,1978AnSta...6..461S,Kass01061995}

\beqa
{\rm AIC} = \chi^{2}_{\rm min} + \frac{2kN}{N-k-1} \\
{\rm BIC} = \chi^{2}_{\rm min} + k{\rm ln}N
\eeqa
where $k$ is the number of parameters in a given model, and $N$ is the total number of data points used in the analysis. 
For model comparison, we calculate the differences between models, e.g., $\Delta \rm AIC/\Delta BIC = AIC/BIC_{\rm model 1} - AIC/BIC_{model 2}$. 
The interpretation of $|\Delta \rm AIC|$ follows established criteria~\cite{SolaPeracaula:2017esw}: 
$|\Delta \rm AIC|<2$: the models are statistically equivalent; 
$2<|\Delta \rm AIC|<6$: the model with the higher AIC has marginally less support; 
$6<|\Delta \rm AIC|<10$: the model with the higher AIC is substantially disfavored; 
$|\Delta \rm AIC|>10$: the model with the higher AIC is strongly ruled out, with compelling evidence favoring the alternative. 
Similar criteria apply to the BIC. 
We calculated $\Delta\rm AIC/\Delta \rm BIC$ between pairs of the four models investigated in this work, and the results 
are shown in Tab.~\ref{tab:aic_bic}. The entries in the table are given in the form $\Delta\rm AIC/\Delta \rm BIC$. 
Each entry corresponds to the AIC/BIC value of the column model minus that of the row model. 

For the $\alpha \Lambda$CDM and $\alpha w$CDM models, compared with the $\Lambda$CDM model, most $\Delta$AIC values indicate 
that these two models are either statistically equivalent or have marginally less support, 
while the $\alpha\Lambda$CDM model is preferred for the CMB+BAO+CC+Union3 dataset. 
The BIC values show that these two models are disfavored by most datasets, except for the CMB+BAO+CC+Union3 dataset. 
For the $\alpha w_0 w_a$CDM model, the $\Delta$AIC values demonstrate that the $\Lambda$CDM model is disfavored by all datasets. 
However, the $\Delta$BIC values show that $\Lambda$CDM is still favored compared with $\alpha w_0 w_a$CDM. 
We note that recent work has highlighted systematic uncertainties in the $w_0 w_a$CDM parametrization arising from early linearization of the nonlinear 
$w(a)$CDM formulation~\cite{Abchouyeh:2025ans_2026}, which may induce systematic $w_0$-$\Omega_{m,0}$ correlations. 
This caveat should be kept in mind when interpreting the marginal AIC/BIC preference for the $\alpha w_0 w_a$CDM model.

For the $J$CDM model, compared with the $\Lambda$CDM model, all $\Delta$AIC and $\Delta$BIC values are positive and larger than 10. 
On the basis of $\Delta$AIC alone, fits to the combined early- and late-time data suggest that $\Lambda$CDM remains competitive with 
the other three models. To interpret this outcome, we refer to Tab.~\ref{tab:sep_parameters_total}, which differentiates fits according to 
early- and late-time data. For instance, the $J$CDM model provides an excellent late-time fit and reduces the $H_0$ tension, but its CMB-only 
fit is strongly inconsistent with Planck, and the combined fit requires a significantly larger $\Omega_bh^2$. Consequently, the large positive 
$\Delta$AIC/$\Delta$BIC for $J$CDM reflects the early-universe cost. Similarly, the AIC preference for $\alpha w_0 w_a$CDM does not translate into 
a higher $H_0$ and therefore does not resolve the Hubble tension. Therefore, a model which can reduce the Hubble tension in 
the late-time fits does not guarantee its overall viability, and the early-universe cost must be quantified.

%%%%%%%%%%%%%%%%%%%%%%%%%%%%%%%%%%%%

\begin{table*}[!htb]
%\footnotesize
\caption{Model comparison with the Akaike Information Criterion (AIC) and the Bayesian Information
Criterion (BIC). Each constraint from the first line to the third line corresponds, respectively, 
to the following datasets: CMB+BAO+CC+PantheonPlus, CMB+BAO+CC+DESY5, and CMB+BAO+CC+Union3. Each entry in the table gives $\Delta$AIC and $\Delta$BIC 
for the column model relative to the row model (i.e., the value for the column model minus that for the row model). 
The values are reported as $\Delta \rm AIC /\Delta BIC$. For example, the values in the first column of the second row are 
$\Delta \rm AIC =  AIC_{\alpha \Lambda CDM} - AIC_{\Lambda CDM}=1.29$ and 
$\Delta \rm BIC =  BIC_{\alpha \Lambda CDM} - BIC_{\Lambda CDM}=6.69$. 
}
\label{tab:aic_bic}
\centering
\begin{ruledtabular} 
\begin{tabular}{lcccccccc}

Model & $\Lambda \rm CDM$ & $\alpha \Lambda \rm CDM$ &$\alpha w\rm CDM$ & $\alpha w_0 w_a\rm CDM$ & $J$CDM \\
\hline
\multirow{3}{*}{\(\Lambda \rm CDM\)}
   & ...   & $-1.29/-6.69$    & $-2.19/-12.98$  & $2.19/-14$    &$-790.76/-790.75$ \\
   & ...   & $-1.75/-7.28$    & $0.69/-10.37$   & $11.94/-4.65$ &$-807.44/-807.43$ \\   
   & ...   & $26.64$/$24.51$  & $-2.67/-6.89$   & $8.29$/$2.07$ &$-749.57/-748.57$ \\  
\hline
\multirow{3}{*}{\(\alpha \Lambda \rm CDM\)}
   & 1.29/6.69           & ...  & $-0.9/-6.29$    & $3.48/-7.31$   &$-789.47/-784.06$ \\
   & 1.75/7.28           & ...  & $2.44/-3.09$    &$13.69$/$2.63$  &$-805.69/-800.15$ \\   
   & $-26.64$/$-24.51$   & ...  & $-29.31/-31.4$  &$-18.35/-22.43$ &$-775.21/-773.07$ \\  
\hline
\multirow{3}{*}{\(\alpha w \rm CDM\)}
   & 2.19/12.98      & 0.9/6.59     & ... & $4.38/-1.02$ &$-788.57/-777.77$ \\
   & $-0.69$/10.37   & $-2.44$/3.09 & ... & $11.25/5.72$ &$-808.13/-797.06$ \\   
   & 2.67/6.89       & 29.31/31.39  & ... & $10.96/8.96$ &$-745.90/-741.68$ \\ 
\hline
\multirow{3}{*}{\(\alpha w_0 w_a \rm CDM\)}
   & $-2.19$/14        & $-3.48$/7.31      & $-4.38/1.02$   & ... &$-792.95/-776.75$ \\
   & $-11.94$/4.65     & $-13.69$/$-2.63$  & $-11.25/-5.71$ &...  &$-819.38/-802.78$ \\   
   & $-8.29$/$-2.07$   & 18.35/22.43       & $-10.96/-8.96$ &...  &$-756.86/-750.64$ \\ 
\hline
\multirow{3}{*}{\(J \rm CDM\)}
   & 790.76/790.75 & 789.47/784.06 &788.57/777.77 &792.95/776.75 &... \\
   & 807.44/807.43 & 805.69/800.15 &808.13/797.06 &819.38/802.78 &...\\
   & 749.57/748.57 & 775.21/773.07 &745.90/741.68 &756.86/750.64 &...\\

\end{tabular}
\end{ruledtabular}
\end{table*}
%%%%%%%%%%%%%%%%%%%%%%%%%%%%%%%%%%%%%

\section{Conclusion}

In this work, motivated by the proposed redshift evolution of the Hubble constant, 
we have investigated four cosmological models. Three of these can be regarded as extensions of the $\Lambda$CDM model 
that incorporate a redshift-dependent Hubble constant, while the fourth is the big bang quantum cosmology $J$CDM model, 
in which the Hubble expansion rate takes a form similar to the other three. Using the datasets CMB+BAO+CC+PantheonPlus/DESY5/Union3, 
we derived constraints on the parameters of these models. For the parameter $\alpha$, which indicates the possible evolution of the Hubble constant, 
it is basically consistent with $\alpha=0$ within $2\sigma$ for all three extended models across all datasets. For the $J$CDM model, 
we found a larger value of the Hubble constant, indicating a reduction in the Hubble tension. We performed AIC and BIC analyses 
for model comparison. We found that the $\Lambda$CDM model remains preferred over the other models. 
The $\alpha w_0 w_a$CDM model shows a marginal AIC advantage ($\rm \Delta AIC \sim -2$ to $-13$) for some datasets, consistent with hints of dynamical dark energy, but this advantage does not translate into a significant increase in $H_0$ and therefore does not resolve the Hubble tension. 
The $J$CDM model, which would provide a possible way to resolve the Hubble tension, 
is ruled out by the present datasets.  

We also conducted the model fits using the early- and late-time data separately, investigating the impact of data from these two epochs on the constraints of the model parameters. For the models $\alpha \Lambda$CDM, $\alpha w$CDM, and $\alpha w_0 w_a$CDM, although the central values of $H_0$ are systematically higher than the late-time values, 
because the CMB-only errors are very large, these differences are not statistically significant. 
The combined results are dominated by the precise late-time data. These results show that the higher $H_0$ for the CMB-only 
fits of these models is caused by the parameter degeneracies, not a physical hint of tension resolution. 
Although the $J$CDM model yields a large $H_0$ which approaches the SH0ES value and reduces the Hubble tension, 
its CMB-only fit gives completely unphysical results, deviating from Planck results significantly. 
Thus, the $J$CDM model alleviates the late-time tension only at the expense of strong inconsistency with early-universe constraints such as $\Omega_bh^2$, 
while the converse holds for $\Lambda$CDM in the face of $H_0$ tension with the Local Distance Ladder.

In terms of relative shifts, the 8.4\% $H_0$ tension is not removed in any of the extended models. For $\alpha\Lambda$CDM, $\alpha w$CDM, 
and $\alpha w_0 w_a$CDM, $H_0$ remains consistent with Planck or is even lower, and the tension is therefore retained. Only $J$CDM reduces the 
late-time $H_0$ tension, but it does so by shifting the discrepancy into $\Omega_bh^2$ at the 7.4\%$-$8.0\% level and into 
$\Omega_mh^2$ at the 4.4\%$-$5.9\% level in the combined fits. This quantifies the early-universe cost of alleviating the late-time tension and 
explains why $J$CDM is strongly disfavored by AIC/BIC despite its good late-time fit.

Overall, our results show that none of the considered models can reconcile the two observational epochs. The combined dataset analysis, while informative, remains insufficient for a thorough assessment of their ability to resolve the Hubble tension, suggesting that either new physics beyond these parametrizations or additional observational probes are needed.

\vspace{1cm}

{\bf Acknowledgements} We are grateful to the anonymous reviewer for their constructive suggestions and comments.

\vspace{1cm}

{\bf Funding} This work is supported by the Shandong Provincial Natural Science Foundation 
(Grant No.ZR2025MS16).

\vspace{1cm}

{\bf Data availability statement} This study employed publicly accessible datasets, which are explicitly cited and referenced in the main text.

\vspace{1cm}

{\bf Competing interests} The authors have no competing interests to declare that are relevant to the content of this article.

\vspace{1cm}

{\bf Authors Contributions} Not applicable.

\bibliographystyle{apsrev4-1}
\bibliography{ref}
\end{document}